\documentclass[12pt]{article}
\usepackage{epsfig,amsfonts,amsthm}
\usepackage{amsmath,amssymb}
\usepackage{cite}
\usepackage{xcolor}
\usepackage{tikz}
\usepackage{extarrows}
\usepackage{graphicx}
\usepackage{siunitx}
\usepackage{caption}
\usepackage{subcaption}
\usepackage{float}

\newcommand{\eqali}[1]{\begin{equation}\begin{aligned}#1\end{aligned}\end{equation}}
\newcommand{\mtrx}[1]{\begin{pmatrix}#1\end{pmatrix}}
\newcommand{\diag}{\operatorname{diag}}

\usepackage{url}

\newcommand{\be}{\begin{equation}}
\newcommand{\ee}{\end{equation}}
\newcommand{\bea}{\begin{eqnarray}}
\newcommand{\eea}{\end{eqnarray}}

\providecommand{\mtrx}[1]{\begin{pmatrix} #1 \end{pmatrix}}

\def\lsim{\mathrel{\rlap{\lower4pt\hbox{\hskip1pt$\sim$}}
    \raise1pt\hbox{$<$}}}         %less than or approx. symbol
\def\gsim{\mathrel{\rlap{\lower4pt\hbox{\hskip1pt$\sim$}}
    \raise1pt\hbox{$>$}}}         %greater than or approx. symbol

\title{Domain walls from $\Sigma(36\times 3)$, $\Delta(54)$ and $\Delta(27)$ potentials}

\author{
Gonçalo~Barreto$^{1}$\thanks{
E-mail: goncalormbarreto@tecnico.ulisboa.pt
},	
Ivo~de~Medeiros~Varzielas$^{1}$\thanks{E-mail: ivo.de@udo.edu},  
    Ye-Ling~Zhou $^{2}$\thanks{E-mail: zhouyeling@ucas.ac.cn}  
	\\
	{\small $^1$ CFTP, 
		Instituto Superior T\'{e}cnico, Universidade de Lisboa,}\\
    \small{ 
		Av. Rovisco Pais 1, 1049-001 Lisboa, Portugal}\\
          {\small $2$ School of Fundamental Physics and Mathematical Sciences,}\\
    \small{
    Hangzhou Institute for Advanced Study, UCAS, Hangzhou 310024, China}
}

\date{}

\begin{document}

\maketitle
    
\begin{abstract}
We consider the degenerate minima arising from scalar potentials invariant under $\Sigma(36\times 3)$, or under its subgroups $\Delta(54)$ and $\Delta(27)$ (with or without imposed CP symmetries), for a triplet of those symmetries. In this framework, we classify the distinct Domain walls between the degenerate minima and calculate the respective tensions.
\end{abstract}

%\tableofcontents

\section{Introduction}

Non-Abelian discrete symmetries provide a powerful framework beyond the Standard Model (SM) for addressing the flavor problem---the mystery of why there are three generations of fermions and why they possess such specific mass hierarchies and mixing patterns. One of the most significant implications is in the lepton sector, where special mixing patterns are predicted and large mixing angles are addressed with the help of non-trivial vacuum expectation values (VEVs) achieved by certain series of scalars \cite{Xing:2020ijf,Feruglio:2019ktm,Ding:2023htn}. 
In multi-Higgs models, non-Abelian discrete symmetries are also motivated in both reducing free parameters in the potential and by interesting possibilities of spontaneous CP violation \cite{Keus:2013hya, Ivanov:2017dad}. 

The list of symmetries for three-Higgs doublet models (3HDMs) is listed in
 \cite{Ivanov:2012fp, Kuncinas:2025uty} \footnote{See also \cite{Darvishi:2019dbh,Darvishi:2021txa}.}, and some recent studies of 3HDMs are performed e.g. in \cite{deMedeirosVarzielas:2019rrp, CarcamoHernandez:2021osw, deMedeirosVarzielas:2021zqs, Kalinowski:2021lvw, deMedeirosVarzielas:2022kbj,  Carrolo:2022oyg, Kuncinas:2023ycz, Barreto:2025jzx, deMedeirosVarzielas:2025byb}. Solving the flavor problem in this framework is discussed  recently in \cite{deMedeirosVarzielas:2019dyu, CarcamoHernandez:2022vjk,Bree:2023ojl, Vien:2024zbj}.

Discrete symmetries in the early Universe have interesting cosmological consequences. 
The spontaneous breaking of discrete symmetry generates Domain walls (DWs), two-dimensional topological defects at cosmic scale \cite{Kibble:1976sj}. They oscillate and radiate gravitational waves,\footnote{In the radiation era, the evolution of DW network follows the scaling behavior, with its energy density diluted much slower than the radiation during the Hubble expansion. To avoid its domination in the Universe, the DW should collapse before the period of big bang nucleosynthesis. This condition is satisfied by including small explicit breaking in the potential.} providing a cosmological probe to discrete symmetries. It was highlighted that stochastic gravitational wave backgrounds from collapsing DWs provide a good fit to data from Pulsar Time Array (PTA) measurement NANOGrav \cite{NANOGrav:2023hvm}. Most studies of DW dynamics start from $Z_2$ DW or axion DW \cite{Saikawa:2017hiv}; see \cite{Kitajima:2023cek,Ferreira:2023jbu,Ferreira:2024eru,Dankovsky:2024zvs,Li:2025gld,Blasi:2025tmn}  for recent updates. In more general potential in larger symmetries, DWs with more complicated properties form and their evolution dynamics can be significantly different from $Z_2$ DW \cite{Wu:2022stu}. Adjacent and nonadjacent walls are specified from the breaking of $Z_N$ symmetries for $N>3$ \cite{Wu:2022tpe}. Properties of non-Abelian DWs were first studied in \cite{Ouahid:2018gpg, Jueid:2023cgp,Fu:2024jhu}, with $A_4$ and $S_4$ as typical examples. Different kinds of DWs are observed due to combinations of rich vacua configurations in $S_4$. Extended studies are further performed in $A_4$ \cite{Fu:2025qhf}. In particular, walls separating two CP-conjugate walls are pointed out when VEVs of complex scalars in $A_4$ are taken into account \cite{Fu:2025qhf}. Gravitational waves from non-Abelian DWs provide a cosmological window to the origin of flavor problem \cite{Gelmini:2020bqg,Chen:2026fod}.
%%%%%%%%%%%%%%%%%%%%%%%%%%%%%%%%%%%%%%%%

In this work, we focus on properties of domain walls formed from $\Delta(27)$, $\Delta(54)$ and $\Sigma(36\times 3)$\footnote{$\Sigma(36 \times 3)$ is also denoted as $\Sigma(36\varphi)$ with $\varphi =3$, and $\Sigma(36)\equiv \Sigma(36 \varphi)/Z_3 \simeq (Z_3 \times Z_3) \rtimes Z_4$ \cite{Fairbairn:1964sga}.}, all of which are subgroups of $SU(3)$ \cite{Grimus:2010ak}, with the following
isomorphism relations satisfied, 
\begin{align}
&\Delta(27) \simeq (Z_3 \times Z_3) \rtimes Z_3\,,\nonumber\\ 
&\Delta(54) \simeq \Delta(27) \rtimes Z_2\,, \nonumber\\
&\Sigma(36\times 3)\simeq \Delta(54) \rtimes Z_2 \simeq \Delta(27) \rtimes Z_4 \,.
\end{align} 
Compared with the most well-studied groups $A_4$ or $S_4$, these groups, due to their complex structures, provide nonrotatable special complex phases in the vacua, giving rise to geometrical CP violation \cite{Branco:1983tn, deMedeirosVarzielas:2011zw, deMedeirosVarzielas:2012ylr, Bhattacharyya:2012pi}.

The layout of the paper is as follows. In Section \ref{sec:pot} we briefly review the potentials invariant under the respective symmetries and their minima. In Section \ref{sec:DW} we classify the distinct Domain walls that can form between the degenerate minima, and calculate the tensions for specific benchmark points in parameter space. We conclude in Section \ref{sec:con}.

\section{$\Sigma(36\times 3)$, $\Delta(54)$ and $\Delta(27)$ potentials \label{sec:pot}}

\subsection{Generators}

It will be convenient in our analysis to consider that $\Delta(27)$ (with 27 elements) is a subgroup of $\Delta(54)$ (with 54 elements), which is a subgroup of $\Sigma(36 \times 3)$ (with 108 elements).

Following the notation in \cite{Barreto:2025jzx}, with $\omega=e^{i 2 \pi/3}$, we take the generators of $\Sigma(36 \times 3)$ for a triplet irreducible representation,

\begin{equation}
a = \mtrx{1&0&0\\ 0&\omega&0\\ 0&0&\omega^2}\,, \quad
b = \mtrx{0&1&0\\ 0&0&1\\ 1&0&0}\,,\quad
d = \frac{i}{\sqrt{3}} \left(\begin{array}{ccc} 1 & 1 & 1 \\ 1 & \omega^2 & \omega \\ 1 & \omega & \omega^2 \end{array}\right)\,.
\label{eq: Sigma36-generators}
\end{equation}

For $\Delta(54)$ we consider similarly the generators,

\begin{equation}
a = \mtrx{1&0&0\\ 0&\omega&0\\ 0&0&\omega^2}\,, \quad
b = \mtrx{0&1&0\\ 0&0&1\\ 1&0&0}\,,\quad
d^2 = \left(\begin{array}{ccc} -1 & 0 & 0 \\ 0 & 0 & -1 \\ 0 & -1 & 0 \end{array}\right)\,.
\label{eq: Delta 54-generators}
\end{equation}

It is the generator $d$ that differentiates between the groups, with $\Delta(54)$ only having the even power $d^2$, and $\Delta(27)$ having just $a$ and $b$ as generators.

The renormalizable scalar potential for a triplet of these groups is similar. When imposing $\Delta(27)$ invariance, the potential is accidentally invariant under generator $d^2$, and thus the potential is the same as in the $\Delta(54)$ case. Throughout the paper we generally refer only to $\Delta(54)$, noting here that everything applies also to the $\Delta(27)$ case.

\subsection{CP symmetries}

The scalar potential invariant under $\Sigma(36 \times 3)$ is known to be invariant under CP, whereas the most general potential invariant under $\Delta(54)$ is not. As noted in \cite{Nishi:2013jqa}, there are a few general CP transformations that do not enlarge the $\Delta(27)$ group. We list them here for convenience.

\eqali{
\label{list:S}
S_0&=\mtrx{1&0&0\cr0&1&0\cr0&0&1}\,,& S_1&=\mtrx{1&0&0\cr0&0&1\cr0&1&0}\,,&
  S_2&=\diag(1,1,\omega)\,,&S_3&=\diag(1,1,\omega^2)\,,
\cr
S_4&=\frac{1}{\sqrt{3}}\mtrx{1&1&1\cr1&\omega&\omega^2\cr1&\omega^2&\omega}\,,& S_5&=S_4 S_1\,,&
  S_6&=\frac{-i\omega}{\sqrt{3}}\mtrx{\omega^2&1&1\cr1&1&\omega^2\cr1&\omega^2&1}\,,&
  S_7&=\frac{i\omega^2}{\sqrt{3}}\mtrx{\omega&1&1\cr1&1&\omega\cr1&\omega&1}\,,
\cr
S_8&=\mtrx{\omega&0&0\cr0&0&1\cr0&1&0}\,,&
  S_9&=\mtrx{\omega^2&0&0\cr0&0&1\cr0&1&0}\,,& 
  S_{10}&=S_6 S_1\,,& S_{11}&=S_7 S_1\,.
}

As mentioned above, the $\Delta(27)$ and $\Delta(54)$ renormalizable potential for one triplet are the same. Here we note further that CP symmetries associated with $S_0$ and with $S_1$ act in the same way in the potential (as the generator $d^2$ being a symmetry of the potential). The same applies for the CP symmetry pairs $S_4$ / $S_5$, $S_6$ / $S_{10}$ and $S_7$ / $S_{11}$.

\subsection{Potentials}

Potentials of triplets of non-Abelian symmetries have simple quadratic terms \cite{deMedeirosVarzielas:2017glw}, which we write as,

\begin{equation}
V = - m^2 (\phi_1^\dagger \phi_1 + \phi_2^\dagger \phi_2 + \phi_3^\dagger \phi_3) + V_4\,.
\end{equation}
The quartic terms depend on the specific symmetry.

In particular, the scalar potential of the 3HDM invariant under $\Sigma(36 \times 3)$ is,
\begin{eqnarray}
V & = &  - m^2 \left[\phi_1^\dagger \phi_1+ \phi_2^\dagger \phi_2+\phi_3^\dagger \phi_3\right]
+ \lambda_1 \left[\phi_1^\dagger \phi_1+ \phi_2^\dagger \phi_2+\phi_3^\dagger \phi_3\right]^2 \nonumber\\
&&
- \lambda_2 \left[|\phi_1^\dagger \phi_2|^2 + |\phi_2^\dagger \phi_3|^2 + |\phi_3^\dagger \phi_1|^2
- (\phi_1^\dagger \phi_1)(\phi_2^\dagger \phi_2) - (\phi_2^\dagger \phi_2)(\phi_3^\dagger \phi_3)
- (\phi_3^\dagger \phi_3)(\phi_1^\dagger \phi_1)\right] \nonumber\\
&&
+ \lambda_3 \left(
|\phi_1^\dagger \phi_2 - \phi_2^\dagger \phi_3|^2 +
|\phi_2^\dagger \phi_3 - \phi_3^\dagger \phi_1|^2 +
|\phi_3^\dagger \phi_1 - \phi_1^\dagger \phi_2	|^2\right)\, .
\label{Vexact}
\end{eqnarray}

The most general scalar potential of the 3HDM invariant under $\Delta(54)$ is the following:

\begin{eqnarray}
V & = &  - m^2 \left[\phi_1^\dagger \phi_1+ \phi_2^\dagger \phi_2+\phi_3^\dagger \phi_3\right]
+ \lambda_1 \left[\phi_1^\dagger \phi_1+ \phi_2^\dagger \phi_2+\phi_3^\dagger \phi_3\right]^2 \nonumber\\
&&
- \lambda_2 \left[|\phi_1^\dagger \phi_2|^2 + |\phi_2^\dagger \phi_3|^2 + |\phi_3^\dagger \phi_1|^2
- (\phi_1^\dagger \phi_1)(\phi_2^\dagger \phi_2) - (\phi_2^\dagger \phi_2)(\phi_3^\dagger \phi_3)
- (\phi_3^\dagger \phi_3)(\phi_1^\dagger \phi_1)\right] \nonumber\\
&&
+ \lambda_3 \left(
|\phi_1^\dagger \phi_2 - \phi_2^\dagger \phi_3|^2 +
|\phi_2^\dagger \phi_3 - \phi_3^\dagger \phi_1|^2 +
|\phi_3^\dagger \phi_1 - \phi_1^\dagger \phi_2	|^2\right) \nonumber\\
&& 
+ \lambda_4 \left(
(\phi_1^\dagger \phi_3) (\phi_2^\dagger \phi_3) +
(\phi_2^\dagger \phi_1) (\phi_3^\dagger \phi_1) +
(\phi_3^\dagger \phi_2) (\phi_1^\dagger \phi_2)	\right) + H.c.\, 
\label{eq: V delta 54 exact}
\end{eqnarray}
Note that both $\Sigma(36 \times 3)$ and $\Delta(54)$ includes subgroup $Z_3 = \{I, \omega I, \omega^2 I\}$. Keeping in mind that the electroweak symmetry has already been satisfied, the real global symmetries to be guaranteed in Eqs.~\eqref{Vexact} and \eqref{eq: V delta 54 exact} are $\Sigma (36) \equiv \Sigma(36\times 3)/Z_3 \simeq (Z_3 \times Z_3) \rtimes Z_4$ and $\Sigma (18) \equiv \Delta(54)/Z_3 \simeq (Z_3 \times Z_3) \rtimes Z_2$, respectively. 

The effect of additionally imposing specific CP symmetries to the $\Delta(54)$ potential is as follows.
$S_0$ or $S_1$ force $\lambda_4$ to be real.
$S_2$ or $S_3$ impose the following relation:
\begin{equation}
\mathrm{Im}(\lambda_4) = \pm \sqrt{3}(\mathrm{Re}(\lambda_4)-\lambda_3)\,,
\label{eq: CP restriction}
\end{equation}
where signs $+$ and $-$ refer to the CP symmetries $S_2$ and $S_3$, respectively.

The other CP symmetries, when imposed on the renormalizable potential, either give rise to the $\Sigma(36 \times 3)$ potential or even to the $SU(3)$ potential.

Although the most general forms of the potentials apply for three $SU(2)$ doublets transforming as triplets of these discrete non-Abelian symmetries, in the following Sections we present for simplicity results corresponding to the similar potentials for $SU(2)$ singlets, where the $\lambda_2$ is not present. The direction of the minima in the triplet direction are the same \cite{Ivanov:2014doa,deMedeirosVarzielas:2017glw}.
The VEVs and the corresponding Domain Wall configurations are electrically neutral.

\subsection{Minima}

The minima for the $\Delta(54)$ potential \cite{Ivanov:2014doa,deMedeirosVarzielas:2017glw} are in the following 4 orbits.
\begin{eqnarray}
\mbox{Alignment $A$:}&& A_1 = (\omega,\,1,\,1)\,, \quad A_2 = (1,\,\omega,\,1),\, \quad A_3 = (1,\,1,\,\omega)\label{points-A} \,. \nonumber\\
\mbox{Alignment $A'$:}&& A'_1 = (\omega^2,\,1,\,1)\,, \quad A'_2 = (1,\,\omega^2,\,1),\, \quad A'_3 = (1,\,1,\,\omega^2)\label{points-Ap} \,. \nonumber\\
\mbox{Alignment $B$:}&& B_1 = (1,\,0,\,0)\,, \quad B_2 = (0,\,1,\,0),\, \quad B_3 = (0,\,0,\,1)\label{points-B} \,. \nonumber\\
\mbox{Alignment $C$:}&& C_1 = (1,\,1,\,1)\,,\quad C_2 = (1,\,\omega,\,\omega^2)\,,\quad C_3 = (1,\,\omega^2,\,\omega) \,. \label{eq: alignments}
\end{eqnarray}

In order to identify what is the residual flavor symmetry subgroup for each of the minima, it is relevant to consider that the potential is invariant under a global rephasing, and to recall that we are considering the breaking of $\Sigma (18) \equiv \Delta(54)/Z_3 \simeq (Z_3 \times Z_3) \rtimes Z_2$.
We list explicitly two elements preserved by one representative for each orbit, namely for $A_1$, $A'_1$, $B_1$, $C_1$. Those two elements will generate the residual subgroup. For the other minima in the same orbit we have elements related by conjugation, and a conjugate residual subgroup. For $A_1$, $A'_1$, $B_1$, $C_1$, we note that $d^2$ (up to the global rephasing by $-1$) is preserved. Additionally,
$A_1$ preserves $a b a$, $A'_1$ preserves $a^2 b a^2$, $B_1$ preserves $a$ and $C_1$ preserves $b$, in all four cases the subgroup generated is isomorphic to $S_3$, but they are different $S_3$ subgroups of the original group. E.g. for explicit illustration, $B_1$ we have the identity $e$, $a$, $a^2$, $d^2$, $a d^2$, $d^2 a$; for $C_1$ we have
$e$, $b$, $b^2$, $d^2$, $b d^2$, $d^2 b$. The isomorphism between two $S_3$ groups is explicit from this, and the generators of the respective subgroups do not commute.
Further, within each orbit, the preserved subgroup of another alignment is necessarily a conjugate subgroup and isomorphic to $S_3$. Explicitly, consider $B_2$ minima, which preserves $b a b^2$ and $b^2 d^2 b$., i.e. the $S_3$ subgroup generated is conjugate by $b$ with respect to the subgroup identified for the $B_1$ minima.

Depending on the parameters of the potential, the minima are:
\begin{eqnarray}
\mbox{$A$:}&& \mathrm{Im}(\lambda_4) < 0\,,\, \sqrt{3} \mathrm{Im}(\lambda_4) < \mathrm{Re}(\lambda_4) - 3 \lambda_3\,,\, \mathrm{Im}(\lambda_4) < \sqrt{3} (\mathrm{Re}(\lambda_4) - \lambda_3) \,.\nonumber\\
\mbox{$A'$:}&& \mathrm{Im}(\lambda_4) > 0\,,\, \sqrt{3} \mathrm{Im}(\lambda_4) > -(\mathrm{Re}(\lambda_4) - 3 \lambda_3)\,,\, \mathrm{Im}(\lambda_4) > -\sqrt{3} (\mathrm{Re}(\lambda_4) - \lambda_3) \,. \nonumber\\
\mbox{$B$:}&&  \sqrt{3} \mathrm{Im}(\lambda_4) > \mathrm{Re}(\lambda_4) - 3 \lambda_3\,,\, \sqrt{3} \mathrm{Im}(\lambda_4) < -(\mathrm{Re}(\lambda_4) - 3 \lambda_3)\,,\, \mathrm{Re}(\lambda_4) > 0 \,. \nonumber\\
\mbox{$C$:}&& \mathrm{Im}(\lambda_4) > \sqrt{3} (\mathrm{Re}(\lambda_4) - \lambda_3)\,,\, \mathrm{Im}(\lambda_4) < -\sqrt{3} (\mathrm{Re}(\lambda_4) - \lambda_3)\,,\, \mathrm{Re}(\lambda_4) < 0 \,. \label{eq: alignment selection Delta 54}
\end{eqnarray}
We obtained these constrains by computing the value of the potential at the minima for each alignment and by comparison with other alignments.

We consider what happens to the minima when the symmetry is enlarged from $\Delta(54)$. For example, when we impose invariance under $d$ (or e.g. the general CP associated with $S_{4}$, $S_5$), we get the $\Sigma(36 \times 3)$ potential. Through the action of $d$ we can see that orbit $B$ merges with orbit $C$. This potential is automatically also CP conserving including the trivial CP ($S_0$), therefore the orbit $A$ merges with $A'$.

When the symmetry is enlarged just with a CP symmetry (without leading to invariance under $d$), we have two distinct cases in terms of orbits.

Imposing e.g. trivial CP ($S_0$), $A$ merges with $A'$ (this happens also for $S_1$).

Imposing either $S_2$ or $S_3$, we have the restriction of eq.(\ref{eq: CP restriction}).
The sign $+$ makes the orbit $A$ merge with orbit $C$, whereas the sign $-$ makes the alignments $A'$ and $C$ merge. So we find that $S_2$ (and $S_8$) merges $A$ with $C$, whereas $S_3$ (and $S_9$) merges $A'$ with $C$.

The other cases lead to invariance under $d$.
The imposition of $S_4$ and $S_5$ symmetries leads to the $\Sigma(36 \times 3)$ scalar potential.

We find also that $CP$ symmetries associated to $S_6$, $S_7$, $S_{10}$, $S_{11}$ connect all orbits, which is understood when verifying they constrain the potential so much that we obtain accidentally the potential corresponding to the continuous $SU(3)$ symmetry.

\section{Domain walls \label{sec:DW}}

We perform an analysis in terms of the symmetry relations between the orbits (sets of minima that are related by symmetry). The minima in these orbits are necessarily degenerate, and during the evolution of the Universe, these potentials can give rise to DWs. Our aim here is to classify the topologically distinct cases.
We do so both by considering the symmetries relating each minima, and by systematically going through each case (in the parametric region where those are the minima) and numerically estimating the tension ($\sigma$).

We created a program that numerically estimates the tension for 3 complex fields (6 degrees of freedom). $\sigma = \int_{-\infty}^{+\infty} \varepsilon \, dz$ , where $\varepsilon$ is the energy density. It numerically minimizes the potentials in eq.(\ref{Vexact}) and eq.(\ref{eq: V delta 54 exact}), and from there the DW profile can be computed numerically employing the method implemented in the CosmoTransitions \cite{Wainwright:2011kj} Python module.

Given that the potentials we are considering are invariant under a global phase, we necessarily had to take this into account, which we have done by varying the phase of one of the two minima and checking the minimum tension. It turns out that the relative phase is relevant to find the minimum tension for the particular cases where two orbits merge, as seen below.

The results now follow for each qualitatively distinct symmetry. Each $\Delta(54)$ case has a corresponding $\Delta(27)$ case as noted previously.

\subsection{$\Delta(54)$ \label{sub:D54}}

\begin{figure}[H]
\centering
\begin{subfigure}[c]{0.25\textwidth}
  \centering
  \includegraphics[width=1.0\linewidth]{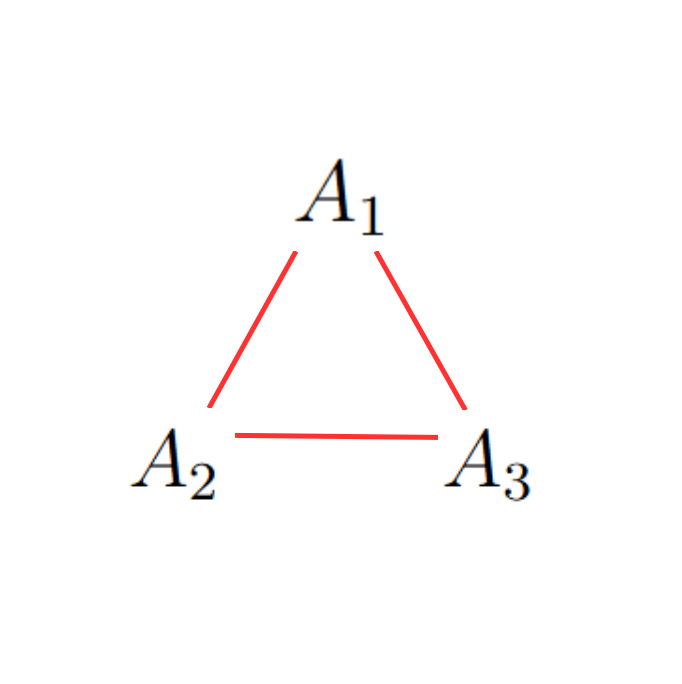}
  \label{fig:sub1}
\end{subfigure}%
\begin{subfigure}[c]{0.25\textwidth}
  \centering
  \includegraphics[width=1.0\linewidth]{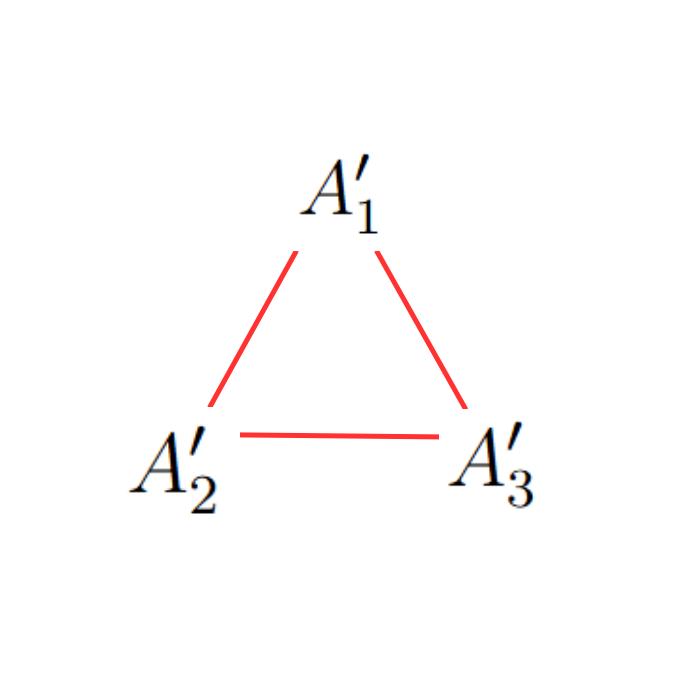}
  \label{fig:sub2}
\end{subfigure}%
\begin{subfigure}[c]{0.25\textwidth}
  \centering
  \includegraphics[width=1.0\linewidth]{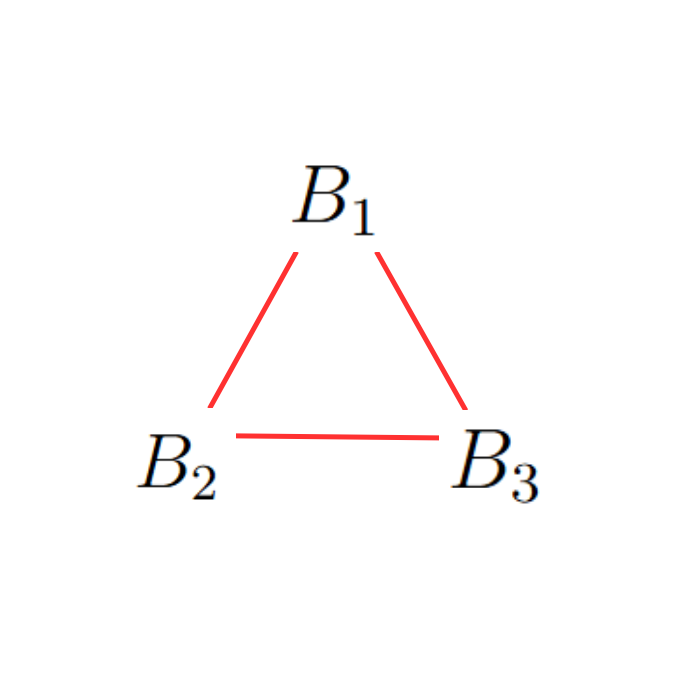}
  \label{fig:sub3}
\end{subfigure}%
\begin{subfigure}[c]{0.25\textwidth}
  \centering
  \includegraphics[width=1.0\linewidth]{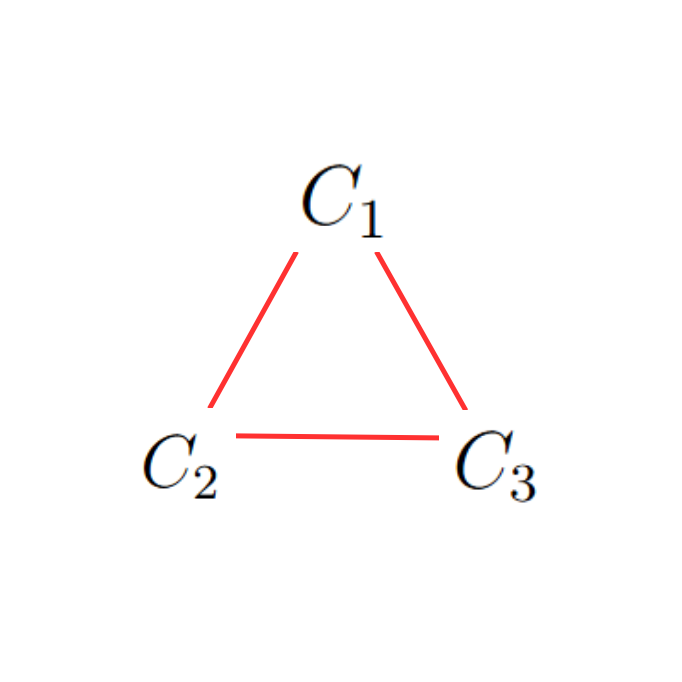}
  \label{fig:sub4}
\end{subfigure}
\caption{Scheme of the DWs in the different alignments for the $\Delta(54)$ without CP symmetries.}
\label{fig: DW schemes Delta54 General}
\end{figure}

The results are summarized in Figure \ref{fig: DW schemes Delta54 General}, where the red lines denote that the minima in that orbit are connected by elements of $\Delta(54)$.

For each case there is only one type of DW. This is confirmed numerically.
To illustrate the results, we present the DW tension ($\sigma$) in Table \ref{ta:D54}, for a representative set of parameters for which the minima associated with each orbit are the true minima.
For all the benchmark points, including those in the following subsections, the parameters were set to $m^2 = 1.0$ and $\lambda_1 = 1.0$.

\begin{table}[H]
\centering
\normalsize
{\footnotesize
    \label{tab: DW table Delta54 General}
    \begin{tabular}{ | c | c | c | c | c |}
    \hline
    Alignment & $\lambda_3$ & $\mathrm{Re}(\lambda_4)$ & $\mathrm{Im}(\lambda_4)$ & $\sigma$ \\ \hline 
    
    $A$ & 0.20 & 0.50 & -0.50 & 0.54 \\ \hline
    
    $A'$ & 0.20 & 0.50 & 0.50 & 0.54 \\ \hline
    
    $B$ & 1.00 & 0.50 & 0.50 & 0.41 \\ \hline
    
    $C$ & 1.00 & -0.50 & 0.50 & 0.75 \\ \hline
    \end{tabular}
    }
     \caption{Table of the DW tensions and parameter values in the different alignments for the $\Delta(54)$ without CP symmetries. \label{ta:D54}}
\end{table}

We note that when reversing the phase of $\lambda_4$ so that either $A$ or $A'$ are the minima, but leaving the remaining parameters unchanged, the numerical value of the tension matches between these two cases, as seen in Table \ref{ta:D54}.

\begin{figure}[H]
\centering

\begin{subfigure}[c]{0.49\textwidth}
  \centering
  \includegraphics[width=\linewidth]{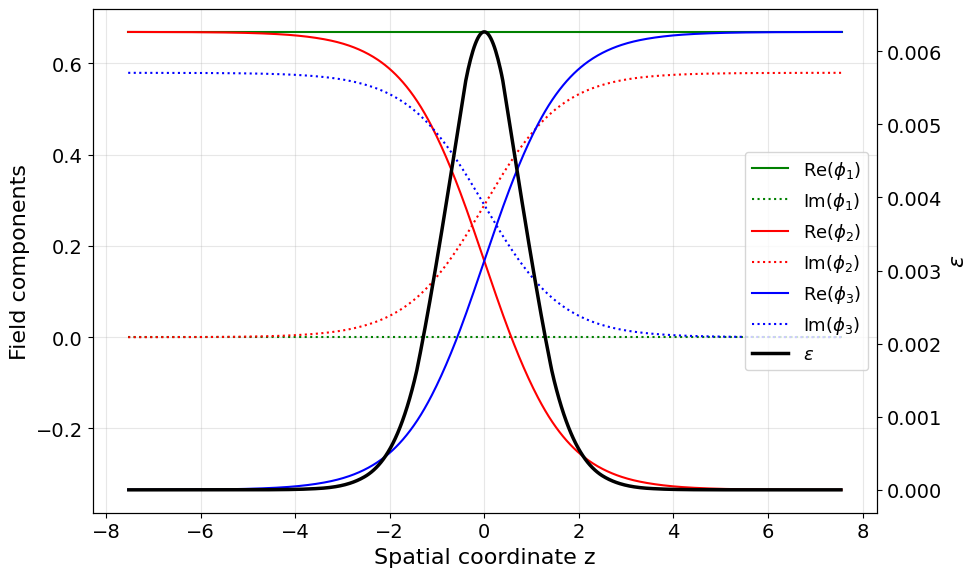}
  \caption{}
  \label{fig:DWA}
\end{subfigure}
\hfill
\begin{subfigure}[c]{0.49\textwidth}
  \centering
  \includegraphics[width=\linewidth]{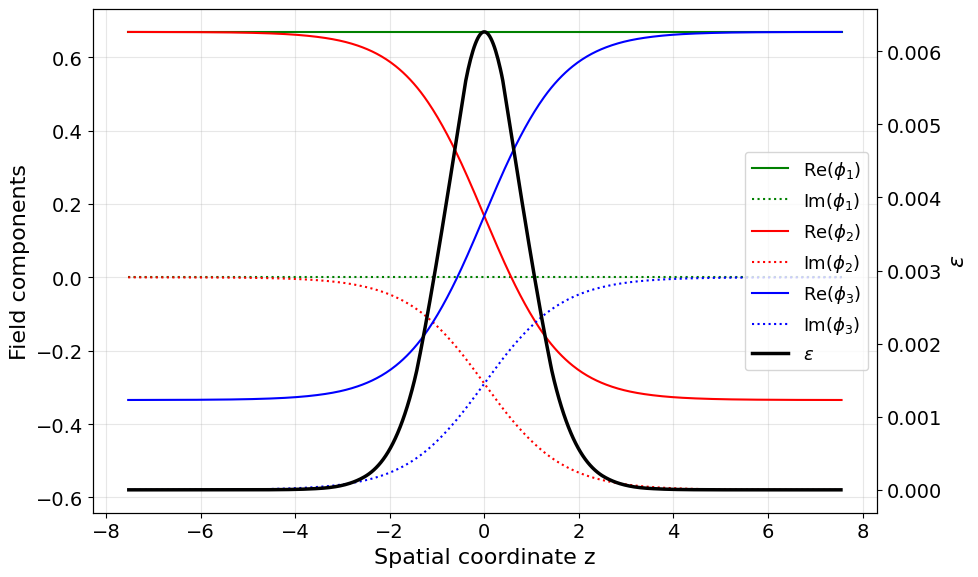}
  \caption{}
  \label{fig:DWA2}
\end{subfigure}

\vspace{0.5cm}

\begin{subfigure}[c]{0.49\textwidth}
  \centering
  \includegraphics[width=\linewidth]{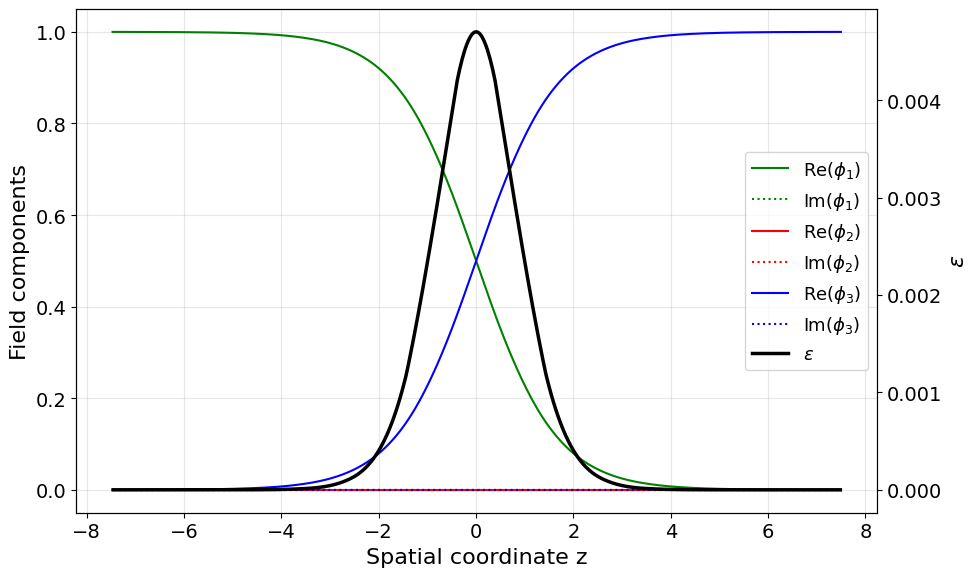}
  \caption{}
  \label{fig:DWB}
\end{subfigure}
\hfill
\begin{subfigure}[c]{0.49\textwidth}
  \centering
  \includegraphics[width=\linewidth]{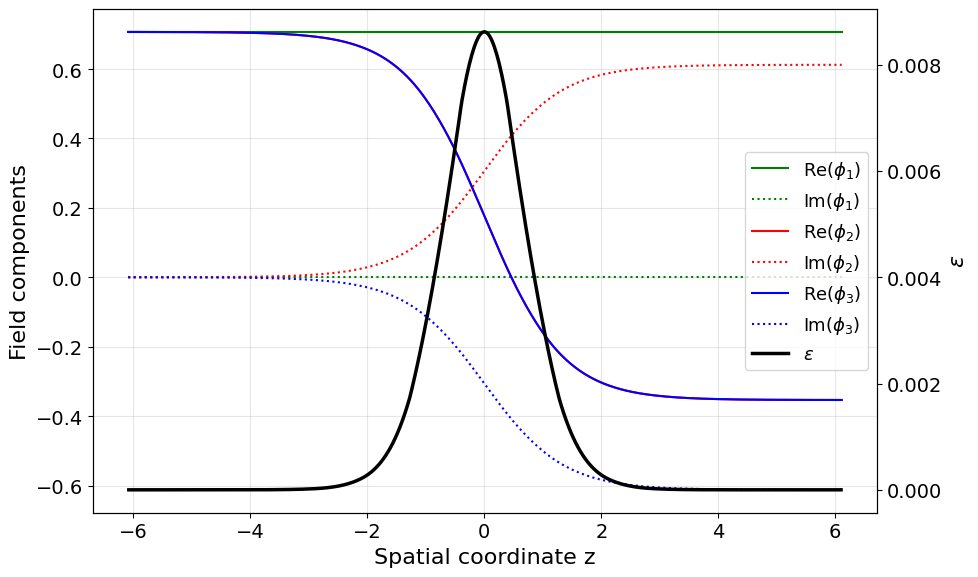}
  \caption{}
  \label{fig:DWC}
\end{subfigure}

\caption{Field profiles with real (solid line) and imaginary (dashed line) parts of the $\phi_i$ in colour, and the energy density (black) as function of spatial coordinate $z$.}
\label{fig:DWschemesDelta54General}
\end{figure}

We display also here, for each of the alignments, the field profiles in terms of spatial coordinate $z$ from $-\infty$ to $+\infty$, as well as the integrand $\varepsilon$. Figs. \ref{fig:DWA}, \ref{fig:DWA2}, \ref{fig:DWB} and \ref{fig:DWC} correspond respectively to $A$, $A'$, $B$ and $C$. As expected from the discussion above, due to the parameters chosen for $A$ and $A'$, Figs. \ref{fig:DWA} and \ref{fig:DWA2} have a visible correspondence.

\subsection{$\Delta(54)$ with trivial CP $S_0$}
\label{sub:D54S0}

%%%%%%%%%%%%%%%%%%%%%%%%%%%%%%%%%%%%%%%%%%%%%%%%%%%%%%%%%%%%%%%%%%%%%%%

\begin{figure}[H]
\centering
\begin{subfigure}[c]{0.5\textwidth}
  \centering
  \includegraphics[width=1.0\linewidth]{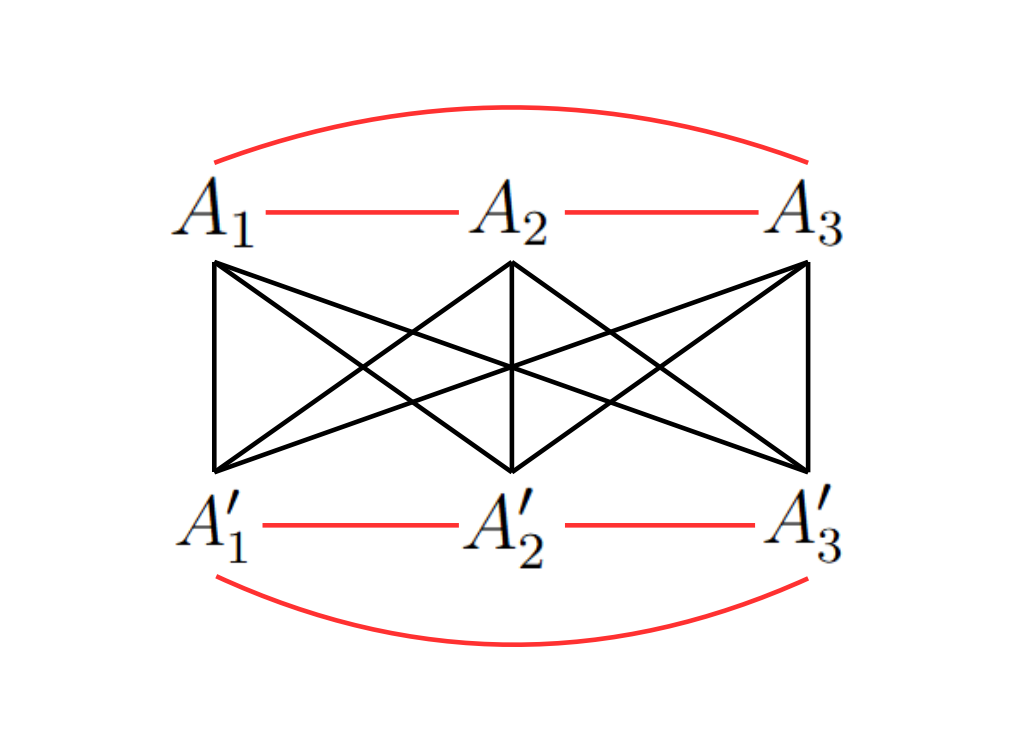}
  \label{fig:sub1}
\end{subfigure}%
\begin{subfigure}[c]{0.25\textwidth}
  \centering
  \includegraphics[width=1.0\linewidth]{Images/AlignmentB.png}
  \label{fig:sub2}
\end{subfigure}%
\begin{subfigure}[c]{0.25\textwidth}
  \centering
  \includegraphics[width=1.0\linewidth]{Images/AlignmentC.png}
  \label{fig:sub1}
\end{subfigure}
\caption{Scheme of the DWs in the different alignments for the $\Delta(54)$ with trivial CP case.}
\label{fig: DW schemes Delta54 S0}
\end{figure}

The $S_0$ CP symmetry, or trivial CP, imposes that $\lambda_4$ is real. Orbit $A$ and $A'$ merge.
The results are summarised in Figure \ref{fig: DW schemes Delta54 S0}.

In contrast with the general case, we find that there are two topologically distinct DWs when the minima are $A$ and $A'$: DWs between two $A_i$ minima are equivalent to DWs between two $A'_i$ minima, shown with red lines denoting their connection through elements of $\Delta(54)$. But the DWs between an $A_i$ and an $A'_j$ minima are distinct, related by a CP symmetry, which we denote through black lines.
When the minima are the other orbits, there is only one type of DW, as in the general case.

Table \ref{ta:S0} has the numerically estimated DW tensions for representative benchmark points.

\begin{table}[H]
\centering
\normalsize
{\footnotesize
    \label{tab: DW table Delta54 S0}
    \begin{tabular}{ | c | c | c | c | c |}
    \hline
    Alignment & $\lambda_3$ & $\mathrm{Re}(\lambda_4)$ & $\mathrm{Im}(\lambda_4)$ & $\sigma$ \\ \hline 
    
    $A$ or $A'$ & -0.10 & 0.10 & 0 & 0.40 \\ \hline
    
    $A + A'$ & -0.10 & 0.10 & 0 & 0.11 \\ \hline
    
    $B$ & 1.00 & 0.30 & 0 & 0.41\\ \hline
    
    $C$ & 1.00 & -0.30 & 0 & 0.57 \\ \hline
    \end{tabular}
    }
     \caption{Table of the DW tensions and parameter values in the different alignments for the $\Delta(54)$ with trivial CP case. \label{ta:S0}}
\end{table} 

\begin{figure}[H]
\centering
\includegraphics[width=0.5\linewidth]{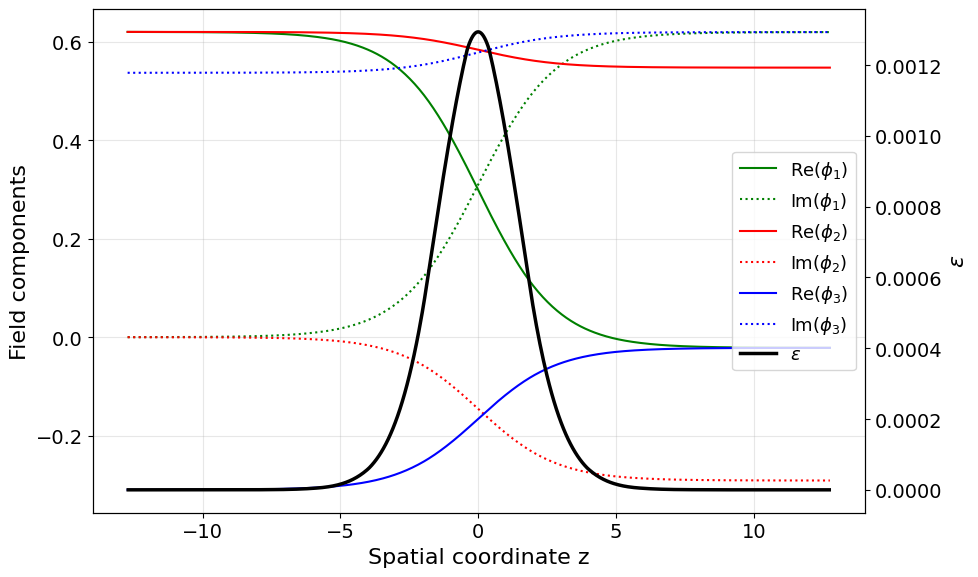}
\caption{Field profiles with real (solid line) and imaginary (dashed line) parts of the $\phi_i$ in colour, and the energy density (black) as function of spatial coordinate $z$. The relative phase between $A_3$ and $A'_2$ is around \ang{92}.}
\label{fig:DWschemesDelta54S0}
\end{figure}

We display in Figure \ref{fig:DWschemesDelta54S0} the field profiles and energy density for the CP DW between alignments from $A$ and $A'$. For the other cases, the profiles are similar to those of Section \ref{sub:D54}.

%%%%%%%%%%%%%%%%%%%%%%%%%%%%%%%%%%%%%%%%%%%%%%%%%%%%%%%%%%%%%%%%%%%%%%%

\subsection{$\Delta(54)$ with $S_2$ CP symmetry}

%%%%%%%%%%%%%%%%%%%%%%%%%%%%%%%%%%%%%%%%%%%%%%%%%%%%%%%%%%%%%%%%%%%%%%%

\begin{figure}[H]
\centering
\begin{subfigure}[c]{0.5\textwidth}
  \centering
  \includegraphics[width=1.0\linewidth]{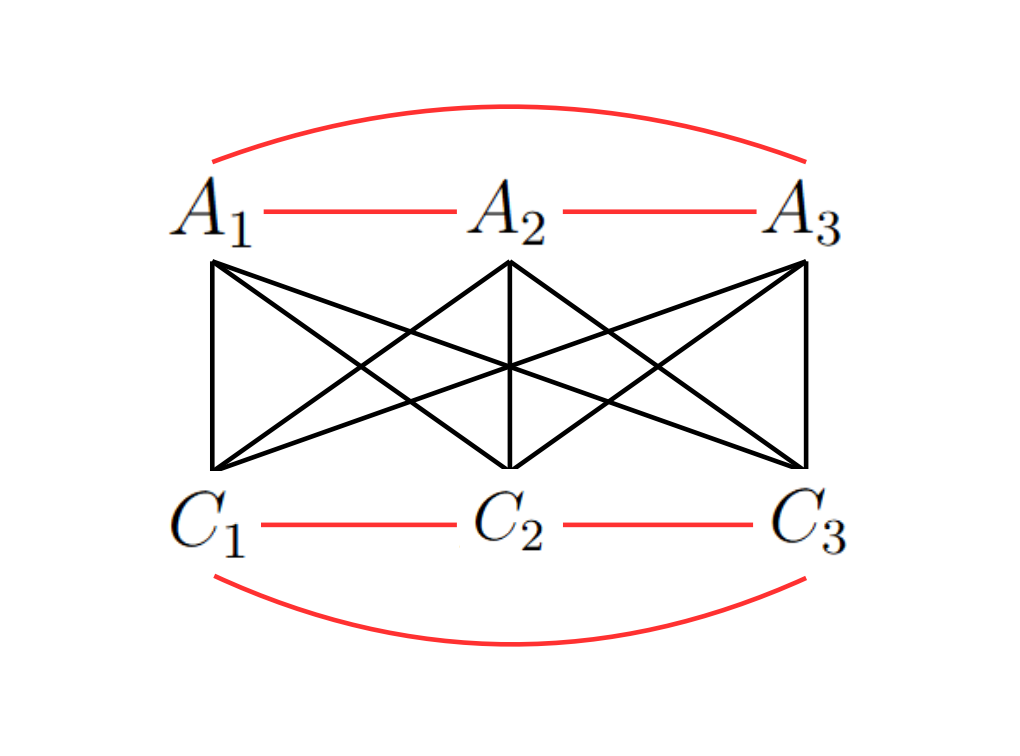}
  \label{fig:sub1}
\end{subfigure}%
\begin{subfigure}[c]{0.25\textwidth}
  \centering
  \includegraphics[width=1.0\linewidth]{Images/AlignmentA2.png}
  \label{fig:sub2}
\end{subfigure}%
\begin{subfigure}[c]{0.25\textwidth}
  \centering
  \includegraphics[width=1.0\linewidth]{Images/AlignmentB.png}
  \label{fig:sub1}
\end{subfigure}
\caption{Scheme of the DWs in the different alignments for the $\Delta(54)$ with $S_2$ CP symmetry.}
\label{fig: DW schemes Delta54 S2}
\end{figure}

The $S_2$ CP symmetry imposes eq.(\ref{eq: CP restriction}) with the $+$ sign. In this case the alignments $A$ and $C$ merge.
The results are summarized in Figure \ref{fig: DW schemes Delta54 S2}.

In contrast with the general case, we find that there are two topologically distinct DWs when the minima are $A$ and $C$, with the DWs between an $A_i$ and a $C_j$ minima, which are related by a CP symmetry (black lines), being distinct.
Table \ref{ta:S2} presents the  numerically estimated DW tensions.

\begin{table}[H]
\centering
\normalsize
{\footnotesize
    \label{tab: DW table Delta54 S2}
    \begin{tabular}{ | c | c | c | c | c |}
    \hline
    Alignment & $\lambda_3$ & $\mathrm{Re}(\lambda_4)$ & $\mathrm{Im}(\lambda_4)$ & $\sigma$ \\ \hline 
    
    $A$ or $C$ & 1.00 & -0.50 & -2.60 & 0.75 \\ \hline
    
    $A + C$ & 1.00 & -0.50 & -2.60 & 0.20 \\ \hline
    
    $A'$ & 0.10 & 0.50 & 0.69 & 0.88 \\ \hline
    
    $B$ & 1.00 & 0.30 & -1.21 & 0.41 \\ \hline
    \end{tabular}
    }
     \caption{Table of the DW tensions and parameter values in the different alignments for the $\Delta(54)$ with $S_2$ CP symmetry. \label{ta:S2}}
\end{table}

\begin{figure}[H]
\centering
\includegraphics[width=0.5\linewidth]{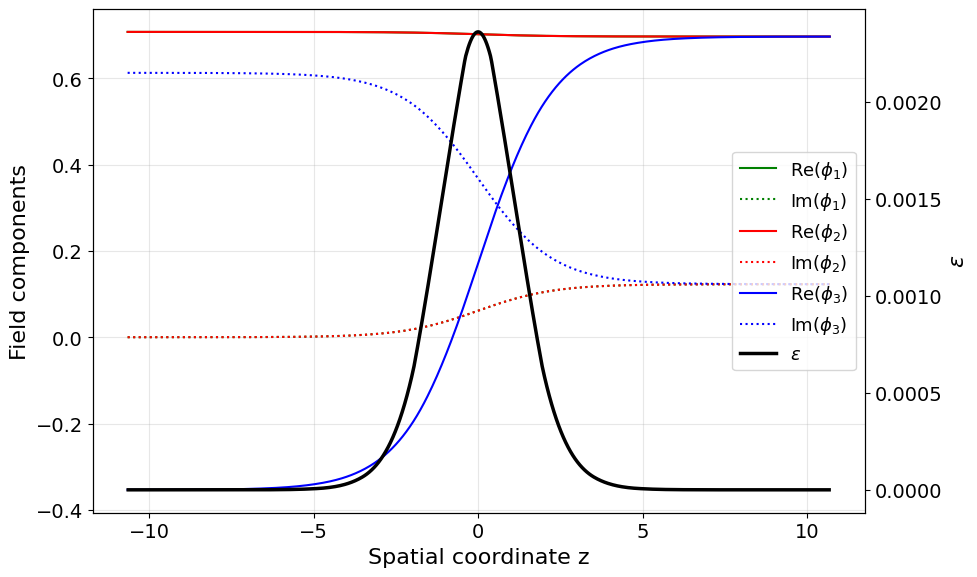}
\caption{Field profiles with real (solid line) and imaginary (dashed line) parts of the $\phi_i$ in colour, and the energy density (black) as function of spatial coordinate $z$. The relative phase between $A_3$ and $C_1$ is around \ang{10}.}
\label{fig:DWschemesDelta54S2}
\end{figure}

We display in Figure \ref{fig:DWschemesDelta54S2} the field profiles and energy density for the CP DW between alignments from $A$ and $C$. For the other cases, the profiles are similar to those of Section \ref{sub:D54}.

%%%%%%%%%%%%%%%%%%%%%%%%%%%%%%%%%%%%%%%%%%%%%%%%%%%%%%%%%%%%%%%%%%%%%%%

\subsection{$\Delta(54)$ with $S_3$ CP symmetry}

%%%%%%%%%%%%%%%%%%%%%%%%%%%%%%%%%%%%%%%%%%%%%%%%%%%%%%%%%%%%%%%%%%%%%%%

\begin{figure}[H]
\centering
\begin{subfigure}[c]{0.25\textwidth}
  \centering
  \includegraphics[width=1.0\linewidth]{Images/AlignmentA.png}
  \label{fig:sub1}
\end{subfigure}%
\begin{subfigure}[c]{0.5\textwidth}
  \centering
  \includegraphics[width=1.0\linewidth]{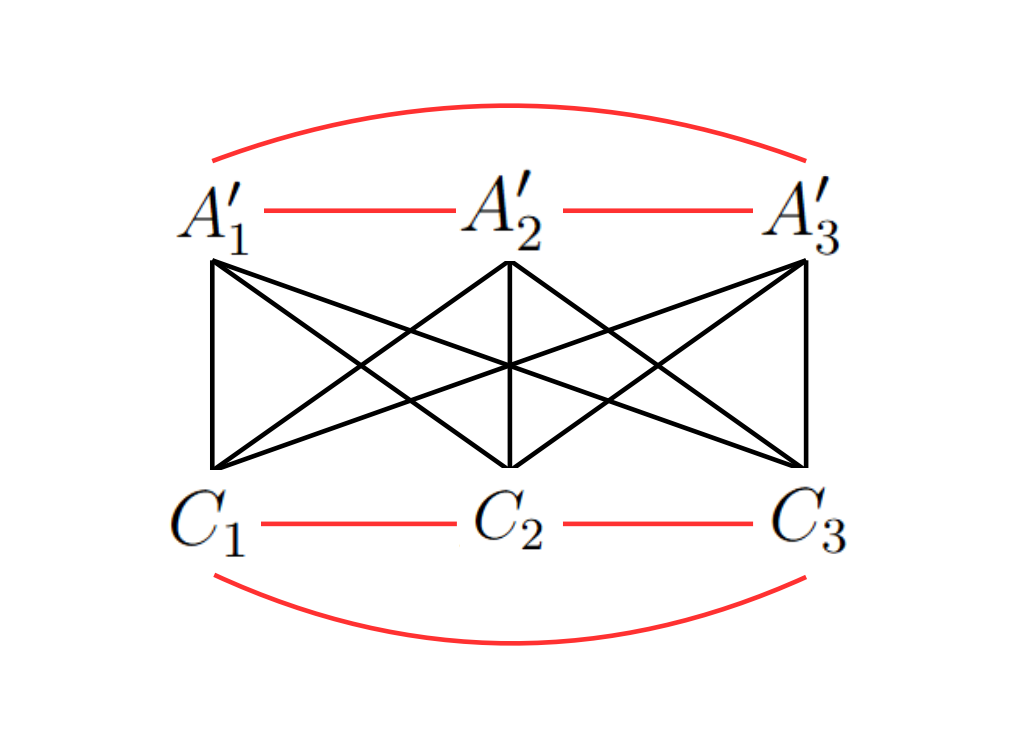}
  \label{fig:sub2}
\end{subfigure}%
\begin{subfigure}[c]{0.25\textwidth}
  \centering
  \includegraphics[width=1.0\linewidth]{Images/AlignmentB.png}
  \label{fig:sub1}
\end{subfigure}
\caption{Scheme of the DWs in the different alignments for the $\Delta(54)$ with $S_3$ CP symmetry.}
\label{fig: DW schemes Delta54 S3}
\end{figure}

The $S_3$ CP symmetry imposes eq.(\ref{eq: CP restriction}) with the $-$ sign. In this case the alignments $A'$ and $C$ merge.
The results are summarised in Figure \ref{fig: DW schemes Delta54 S3}.

In contrast with the general case, and similarly to the $S_2$ case, we find that there are two topologically distinct DWs when the minima are $A'$ and $C$, with the DWs between an $A'_i$ and a $C_j$ minima, which are related by a CP symmetry (black lines) being distinct.
Table \ref{ta:S3} presents the  numerically estimated DW tensions.

\begin{table}[H]
\centering
\normalsize
{\footnotesize
    \label{tab: DW table Delta54 S3}
    \begin{tabular}{ | c | c | c | c | c |}
    \hline
    Alignment & $\lambda_3$ & $\mathrm{Re}(\lambda_4)$ & $\mathrm{Im}(\lambda_4)$ & $\sigma$ \\ \hline 

    $A$ & 0.10 & 0.50 & -0.69 & 0.88 \\ \hline
    
    $A'$ or $C$ & 1.00 & -0.50 & 2.60 & 0.75 \\ \hline
    
    $A' + C$ & 1.00 & -0.50 & 2.60 & 0.20 \\ \hline
    
    $B$ & 1.00 & 0.30 & 1.21 & 0.41 \\ \hline
    \end{tabular}
    }
     \caption{Table of the DW tensions and parameter values in the different alignments for the $\Delta(54)$ with $S_3$ CP symmetry. \label{ta:S3}}
\end{table}

We had found in Table \ref{ta:D54} the same tensions for $A$ and $A'$, as we had selecting corresponding points in parameter space, namely by reversing only the phase of $\lambda_4$. There is a similar correspondence when comparing the tensions for the CP symmetries $S_2$ and $S_3$. For this reason, Tables \ref{ta:S2} and \ref{ta:S3} have corresponding parameters.
The resulting tension for the alignment $B$ is the same in both Tables, and the tension for the unmerged orbit $A'$ of $S_2$ is the same as the tension for the unmerged orbit $A$. The tension of the merged orbits for $S_2$ when connected by symmetries that are CP symmetries ($A + C$) or not ($A$ or $C$) are respectively the same as the tensions of the merged orbits for $S_3$ when connected by symmetries that are CP symmetries ($A' + C$) or not ($A'$ or $C$).

\begin{figure}[H]
\centering
\includegraphics[width=0.5\linewidth]{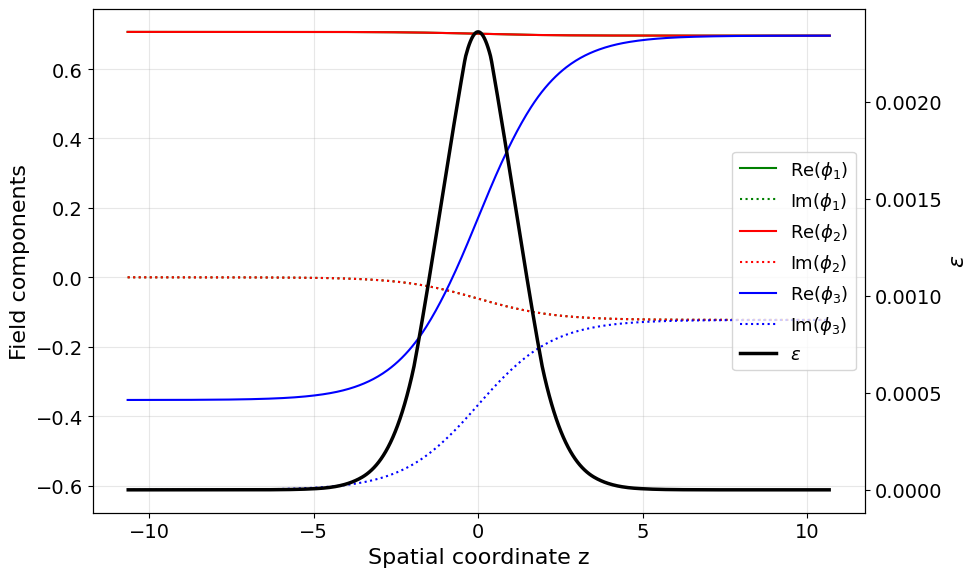}
\caption{Field profiles with real (solid line) and imaginary (dashed line) parts of the $\phi_i$ in colour, and the energy density (black) as function of spatial coordinate $z$. The relative phase between $A_3'$ and $C_1$ is around \ang{350}.}
\label{fig:DWschemesDelta54S3}
\end{figure}

We display in Figure \ref{fig:DWschemesDelta54S3} the field profiles and energy density for the CP DW between alignments from $A'$ and $C$. We note (as expected from the discussion above) the similarity with the results obtained for the CP DW between alignments from $A$ and $C$.

%%%%%%%%%%%%%%%%%%%%%%%%%%%%%%%%%%%%%%%%%%%%%%%%%%%%%%%%%%%%%%%%%%%%%%%

\subsection{$\Sigma(36 \times 3)$}

\begin{figure}[H]
\centering
\begin{subfigure}[c]{0.5\textwidth}
  \centering
  \includegraphics[width=1.0\linewidth]{Images/AlignmentA+A2.png}
\end{subfigure}%
\begin{subfigure}[c]{0.5\textwidth}
  \centering
  \includegraphics[width=1.0\linewidth]{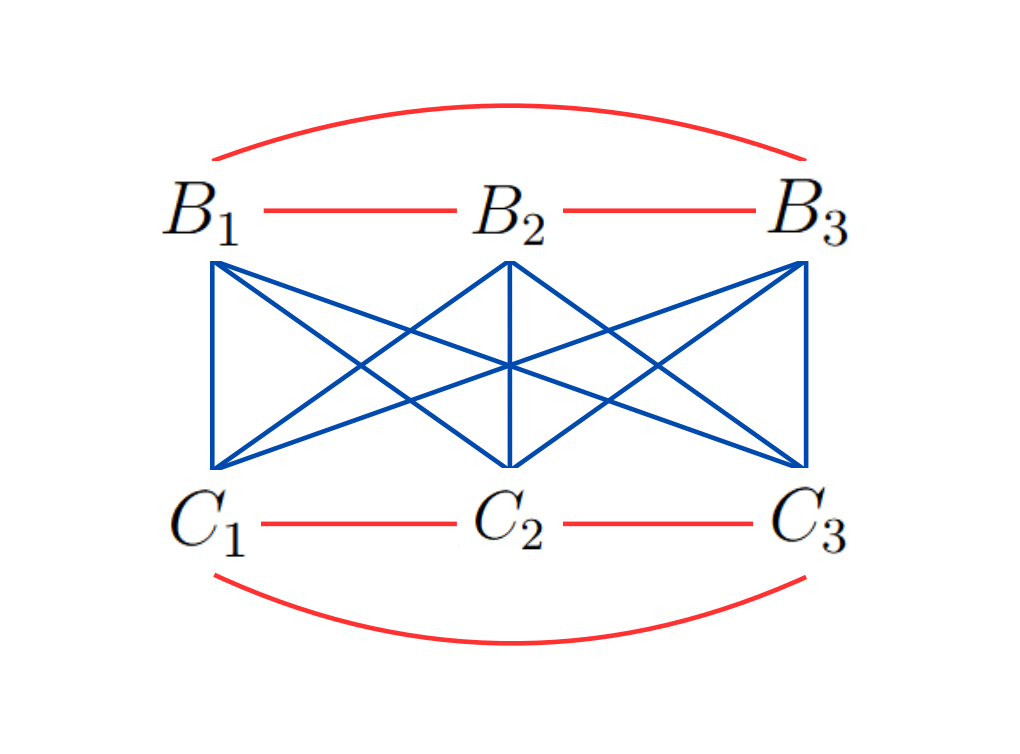}
\end{subfigure}
\caption{Scheme of the DWs in the different alignments for the $\Sigma(36 \times 3)$ case.}
\label{fig: DW schemes Sigma36}
\end{figure}

In the $\Sigma(36 \times 3)$ case, the potential is invariant under the $d$ generator and under CP symmetries, including the trivial CP symmetry $S_0$. The orbit $A$ merges with $A'$ (by action of a CP symmetry) and the orbit $B$ merges with orbit $C$ (by action of $d$).
The results are summarised in Figure \ref{fig: DW schemes Sigma36}.

As expected, the scheme for alignments $A$ and $A'$ are the same as those in Figure \ref{fig: DW schemes Delta54 S0}: the $\Sigma(36 \times 3)$ potential is a specific case of the $\Delta(54)$ potential (with $\lambda_4 = 0$).

A qualitatively different result appears when the minima are $B$ and $C$, where we find two distinct types of DWs.
DWs between two $B_i$ minima are equivalent to DWs between two $C_i$ minima, related by elements of $\Delta(54)$ (red lines). DWs between a $B_i$ and a $C_j$ minima are distinct, and are related by $d$ (or another element of $\Sigma(36 \times 3)$ which is not part of $\Delta(54)$), denoted by blue lines.

Table \ref{ta:S36} has the numerically estimated DW tensions for representative benchmark points.

\begin{table}[H]
\centering
\normalsize
{\footnotesize
    \label{tab: DW table Sigma36}
    \begin{tabular}{ | c | c | c | c | c |}
    \hline
    Alignment & $\lambda_3$ & $\mathrm{Re}(\lambda_4)$ & $\mathrm{Im}(\lambda_4)$ & $\sigma$ \\ \hline 
    
    $A$ or $A'$ & -0.10 & 0 & 0 & 0.38 \\ \hline
    
    $A + A'$ & -0.10 & 0 & 0 & 0.11 \\ \hline
    
    $B$ or $C$ & 1.00 & 0 & 0 & 0.41 \\ \hline
    
    $B + C$ & 1.00 & 0 & 0 & 0.13 \\ \hline
    \end{tabular}
    }
     \caption{Table of the DW tensions and parameter values in the different alignments for the $\Sigma(36 \times 3)$ case. \label{ta:S36}}
\end{table}

\begin{figure}[H]
\centering
\includegraphics[width=0.5\linewidth]{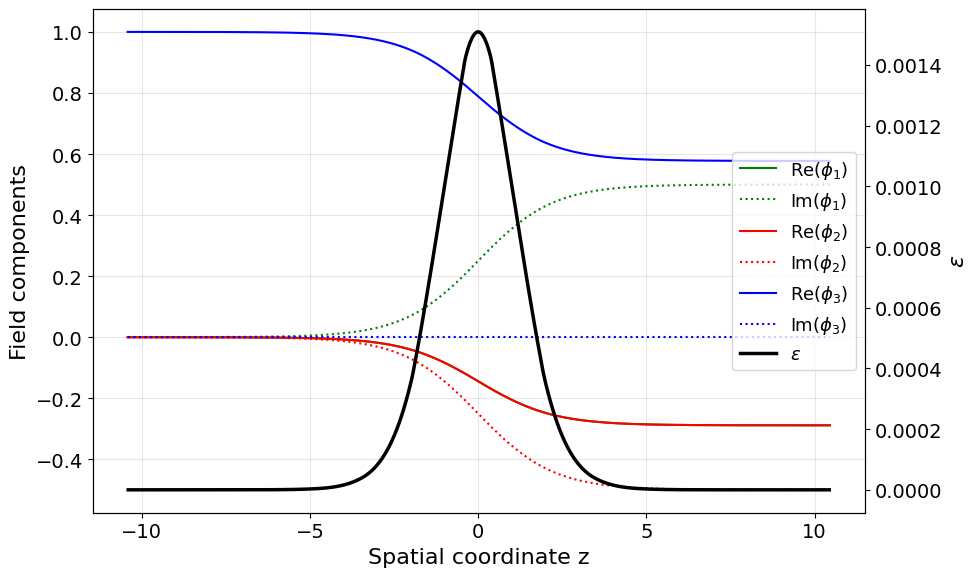}
\caption{Field profiles with real (solid line) and imaginary (dashed line) parts of the $\phi_i$ in colour, and the energy density (black) as function of spatial coordinate $z$. The relative phase between $B_3$ and $C_2$ is around \ang{120}.}
\label{fig:DWschemesSigma36}
\end{figure}

We display in Figure \ref{fig:DWschemesSigma36} the field profiles and energy density for the DW between alignments from $B$ and $C$. For the DW between alignments from $A$ and $A'$, the profiles are similar to those found for trivial CP $S_0$, as expected. For the other cases, the profiles are similar to those discussed in Sections \ref{sub:D54} and \ref{sub:D54S0}.

%%%%%%%%%%%%%%%%%%%%%%%%%%%%%%%%%%%%%%%%%%%%%%%%%%%%%%%%%%%%%%%%%%%%%%%

\section{Conclusion}\label{sec:con}

We review the scalar potentials invariant under symmetries $\Sigma(36 \times 3)$ and its subgroup $\Delta(54)$. The minima of these potentials falls into orbits: sets of degenerate minima, related to each other by symmetries of the potential.

We analyse the possible Domain walls that could form between degenerate minima within each orbit, considering which type of symmetry relates each pair of minima. We also calculate numerically the Domain Wall tensions, which we provide for benchmark points in the parameter space of the potential, where the minima belonging to each orbit are the correct minima.

The classification according to the symmetry relations between minima and the numerical calculation agree.
In the most general case we have $\Delta(54)$ without any CP symmetry. There are four orbits we label $A, A', B, C$, each with their respective Domain Wall type and tension.

Adding different CP symmetries to $\Delta(54)$ will merge two of the 4 general orbits. For example, the trivial CP and related CP symmetries merge $A$ and $A'$ into a single orbit. Other CP symmetries merge $A$ with $C$; and yet other CP symmetries merge $A'$ with $C$. In these cases, we find that the tensions of the previously separate orbits become the same, but there are distinct tensions for Domain walls connecting the (now degenerate) minima of what were previously separate orbits.
Taking the trivial CP case: the tension between $A_i$ and $A_j$ becomes equal to the tension between $A'_i$ and $A'_j$; there is a different tension between $A_i$ and $A'_j$.

In the most symmetric case, we have $\Sigma(36 \times 3)$, and the potential is automatically invariant under CP symmetries such as the trivial CP, which merges  $A$ and $A'$, leading to the same structure as above. Furthermore, the elements of $\Sigma(36 \times 3)$ that are not elements of $\Delta(54)$ merge the orbits $B$ and $C$. We find that the tension between $B_i$ and $B_j$ becomes equal to the tension between $C_i$ and $C_j$, and that there is a different tension between $B_i$ and $C_j$. 

%%%%%%%%%%%%%%%%%%%%%%%%%%%%%%%%%%%%%%%%%%%%%%%%%%%%%%%%%%%%%%%%%%%%%%

\section*{Acknowledgments}
GB thanks Pedro N. de Figueiredo for insightful discussions during the early stages of developing the program used in this work.
IdMV thanks the University of Basel for hospitality.
IdMV acknowledges funding from Fundação para a Ciência e a Tecnologia (FCT) through the FCT Mobility program, and through
the projects CFTP-FCT Unit 
UID/00777/2025 (\url{https://doi.org/10.54499/UID/00777/2025}),  UIDB/FIS/00777/2020 and UIDP/FIS/00777/2020, CERN/FIS-PAR/0019/2021,
CERN/FIS-PAR/0002/2021, 2024.02004 CERN, which are partially funded through POCTI (FEDER), COMPETE,
QREN and EU. 
YLZ was partially supported by National Natural Science Foundation of China (NSFC) under Grant Nos. 12535007, 12547104, and Zhejiang Provincial Natural Science Foundation of China under Grant No. LDQ24A050002. 

%%%%%%%%%%%%%%%%%%%%%%%%%%%%%%%%%%%%%%%%%%%%%%%%%%%%%%%%%%%%%%%%%%%%%%

\bibliographystyle{unsrt}

\end{document}